\documentclass[aps,letterpaper,twocolumn,prl,showpacs]{revtex4-1}
\usepackage{amssymb}
\usepackage{amsmath,bm}

\usepackage[pdftex]{graphicx,color}
\usepackage[english]{babel}
\usepackage{hyperref}
\usepackage{epstopdf}
\begin{document}

\title{Impact of de-synchronization and drift on soliton-based Kerr frequency combs in the presence of pulsed driving fields}

\author{Ian Hendry$^{1,2}$}
\email{i.hendry@auckland.ac.nz}
\author{Bruno Garbin$^{1,2}$}
\altaffiliation{Current address: Centre de Nanosciences et de Nanotechnologies (C2N), CNRS - Universit\'e Paris-Sud - Universit\'e Paris-Saclay, Palaiseau, France}
\author{Stuart G. Murdoch$^{1,2}$}
\author{St\'ephane Coen$^{1,2}$}
\author{Miro Erkintalo$^{1,2}$}
\email{m.erkintalo@auckland.ac.nz}

\affiliation{$^1$The Department of Physics, The University of Auckland, Auckland 1010, New Zealand}
\affiliation{$^2$The Dodd-Walls Centre for Photonic and Quantum Technologies, New Zealand}

\begin{abstract}
Pulsed driving of Kerr microresonators represents a promising avenue for the efficient generation of soliton states associated with coherent optical frequency combs. The underlying physics has not, however, yet been comprehensively investigated. Here, we report on a numerical and theoretical study of the impact of de-synchronization between the periodic pump field and the train of solitons circulating in the cavity. We show that de-synchronization can affect the soliton configurations that can be sustained for given parameters, and that it can be leveraged to guarantee operation in the attractive single-soliton regime. We also reveal that the interplay between pump-resonator de-synchronization and stimulated Raman scattering can give rise to rich dynamics that explain salient features observed in recent experiments. Our work elucidates the dynamics of Kerr cavity solitons in the presence of pulsed driving fields, and could facilitate the development of efficient microresonator frequency comb systems.
\end{abstract}
\maketitle
\section{Introduction}

The generation of optical frequency combs in Kerr nonlinear microresonators has attracted growing interest over the past decade~\cite{del'haye_optical_2007,kippenberg_microresonator-based_2011,pasquazi_microcombs_2018}. Of particular significance has been the discovery that these systems can support ultra-short pulses known as dissipative temporal Kerr cavity solitons (CSs)~\cite{leo_temporal_2010,herr_temporal_2014}, which correspond to coherent and broadband frequency combs in the spectral domain~\cite{kippenberg_dissipative_2018}. Thanks to the many potential applications of soliton-based Kerr frequency combs~\cite{suh_microresonator_2016,marin-palomo_microresonator-based_2017,trocha_ultrafast_2018, suh_soliton_2018, dutt_on-chip_2018,obrzud_microphotonic_2019}, there is continuous interest in improving our understanding of CS behaviour under a variety of operating conditions~\cite{leo_dynamics_2013, hansson_frequency_2015, milian_solitons_2015, yi_theory_2016, anderson_observations_2016, brasch_photonic_2016, anderson_coexistence_2017, xue_second-harmonic_2016,  cole_soliton_2017, jang_synchronization_2018, cole_kerr_2018, wang_stimulated_2018}.

The vast majority of Kerr comb systems employ continuous wave (CW) driving and resonators that exhibit anomalous dispersion. In these systems, CSs manifest themselves as bright pulses of light that sit atop a CW background~\cite{leo_temporal_2010}. Whilst CW driving in principle offers the advantage of reduced system complexity, the small temporal overlap between the CW background and the soliton typically results in very small pump-to-comb conversion efficiency~\cite{bao_nonlinear_2014}. Moreover, precise control of the frequency and power of the driving laser is generally required to reach the desired soliton states~\cite{herr_temporal_2014, brasch_photonic_2016}. Some of these disadvantages can potentially be mitigated by leveraging dark pulse dynamics in normal dispersion resonators~\cite{xue_mode-locked_2015}, or by utilizing systems composed of two coupled resonators~\cite{xue_super-efficient_2019}.

Another solution that has recently attracted particular attention replaces the CW driving field with a train of short pulses whose repetition rate is (approximately) synchronized to the cavity round trip time~\cite{obrzud_temporal_2017, obrzud_microphotonic_2019}. Pioneering experiments have indeed shown that pulsed driving can enable deterministic and controlled generation of CS combs at average power levels significantly lower than those required under conditions of CW driving~\cite{obrzud_temporal_2017}. However, in order to fully take advantage of such pulsed driving schemes, it is important to develop a comprehensive understanding of CS physics in the presence of driving fields with inhomogeneous amplitude profiles. While the dynamics and behaviour of CSs in the presence of driving field \emph{phase} inhomogeneities has been extensively studied~\cite{firth_optical_1996,maggipinto_cavity_2000,jang_temporal_2015,anderson_observations_2016,barland_temporal_2017}, the impact of amplitude inhomogeneities has received comparatively less attention.

We have recently demonstrated that CSs in Kerr resonators driven with pulsed or amplitude modulated fields are attracted to (and subsequently pinned to) temporal positions associated with specific \emph{values} of the driving field amplitude~\cite{hendry_spontaneous_2018} --- behaviour that is in stark contrast to the tendency of CSs to be attracted to the extrema of \emph{phase} inhomogeneities~\cite{maggipinto_cavity_2000,jang_temporal_2015,cole_kerr_2018}. Moreover, Tabbert et al. have shown that amplitude inhomogeneities can affect the domains of soliton existence and stability~\cite{tabbert_stabilization_2019}. The analyses reported in Refs.~\cite{hendry_spontaneous_2018} and~\cite{tabbert_stabilization_2019} were, however, performed under the assumption of perfect synchronization between the CS(s) and the periodic driving field, which is unlikely to hold true in general.

While it is well known that a mismatch between the pump pulse repetition rate and the cavity round trip time can have significant impact on the bistability dynamics of driven passive Kerr resonators~\cite{coen_convection_1999}, the influence of such de-synchronization on CSs has not yet been extensively studied. Parra-Rivas et al. considered the impact of de-synchronization when the driving field is composed of a small intensity perturbation (shorter than the CSs) atop a CW driving field, demonstrating that the resulting temporal drift can affect the stability of CSs~\cite{parra-rivas_effects_2014}. These findings cannot, however, be immediately translated to situations pertinent to recent Kerr comb experiments~\cite{obrzud_temporal_2017,obrzud_microphotonic_2019}, where the driving field comprises of temporally localized pulses longer than the CSs with no CW background. Preliminary simulations~\cite{hendry_dynamics_2018, chen_temporal_2018} considering purely localized driving pulses show that, in accordance with the general behaviour of dissipative solitons in the presence of convective drift~\cite{jang_temporal_2015, cole_kerr_2018, barland_temporal_2017, javaloyes_dynamics_2016, camelin_electrical_2016, garbin_interaction_2017}, CSs can remain (frequency) locked to the driving pulse train over a finite range of de-synchronization, but their trapping positions are shifted compared to the situation of perfect synchronization. Because of the significant application potential of soliton frequency combs generated in microresonators driven by optical pulses~\cite{obrzud_temporal_2017}, there is clearly a need for more detailed understanding of the impact of de-synchronization on the CS dynamics.

In this Article, we report on a systematic theoretical and numerical study of the effect of de-synchronization on CSs in Kerr resonators driven with short pulses. We show that, for typical driving pulse profiles, de-synchronization gives rise to asymmetric shifts of the CS trapping positions, which can influence the multiplicity of possible soliton configurations that the system can support. In particular, we show that de-synchronization can be leveraged to guarantee operation in the single-soliton regime that is most attractive for practical Kerr comb applications. We also investigate the interplay between stimulated Raman scattering and pump-resonator de-synchronization, finding evidence of rich dynamics that explain CS behaviours observed --- but not fully explained --- in recent experiments~\cite{obrzud_temporal_2017}. In addition to elucidating the dynamics of CSs under conditions of localized driving fields, our work could have impact on the design of efficient Kerr comb systems~\cite{obrzud_temporal_2017,obrzud_microphotonic_2019}. Our findings may also be relevant to other Kerr resonator systems where pulsed driving is used to support CSs, such as fiber ring resonators~\cite{anderson_observations_2016, anderson_coexistence_2017, wang_stimulated_2018} or free-space enhancement cavities~\cite{lilienfein_temporal_2019}.

\section{Basic model}

We consider a Kerr resonator with anomalous dispersion that is driven with a train of pulses. Under suitable parameter conditions, a short CS can form atop the driving pulse envelope, as experimentally demonstrated in~\cite{obrzud_temporal_2017}. We are interested in situations where the period of the driving pulse train, $t_\mathrm{P}$, is different from the intrinsic cavity round trip time, $t_\mathrm{R} = \text{FSR}^{-1}$, where $\text{FSR} = v_\mathrm{g}/L$ is the free-spectral range with $L$ the round trip length of the cavity and $v_\mathrm{g}$ the group velocity at the driving wavelength. We model this system using a dimensionless Lugiato-Lefever equation that includes a convective drift term~\cite{coen_convection_1999}:
\begin{equation}
    \label{LLE}
  \frac{\partial E(t,\tau)}{\partial t} = \left[ -1 +i(|E|^2- \Delta)-d\frac{\partial}{\partial\tau}
   +i\frac{\partial^2}{\partial\tau^2}\right]E+S(\tau).
\end{equation}
Here, $t$ is a slow time variable that describes the evolution of the slowly-varying intra-cavity field envelope $E(t,\tau)$ at the scale of the cavity photon lifetime, while $\tau$ is a corresponding fast time that describes the envelope's temporal profile over a single round trip. The terms on the right-hand side of ~Eq.~\eqref{LLE} describe, respectively, the cavity losses, the Kerr nonlinearity, the cavity phase detuning ($\Delta$ being the cavity detuning coefficient), the relative temporal drift between intracavity field and the driving pulse ($d$ being the drift coefficient; see below for normalization), the anomalous group-velocity dispersion, and the fast time dependent coherent driving. In this work, we consider a Gaussian driving field profile, $S(\tau) = S_\mathrm{0}\exp\left[-\tau^{2}/(2\tau_\mathrm{g}^{2})\right]$, where $\tau_{g}$ and $S_\mathrm{0}$ represent the duration and amplitude of the driving pulse, respectively. In line with recent experiments, we focus on the situation where the duration of the driving pulses are (much) longer than the duration of the CSs ($\tau_\mathrm{g} > \Delta^{-1/2}$). We must emphasize, however, that our general findings are not restricted to any particular driving field profile or set of parameters, provided that the driving pulses are (much) longer than the CSs.

Equation~\eqref{LLE} is expressed in a reference frame where the driving pulse is stationary. As a consequence, de-synchronization manifests itself as a constant time-domain drift of intracavity features that are not (fully or partially) locked to the driving pulse. The coefficient $d$ describes the drift in fast time per unit slow time, and is related to the corresponding dimensional parameter \mbox{$\Delta t = t_\mathrm{R} - t_\mathrm{P}$} through the normalization \mbox{$d = \Delta t \sqrt{2\mathcal{F}/(|\beta_2|L\pi)}$}, where $\mathcal{F}$ is the finesse of the cavity and $\beta_2$ is the group-velocity dispersion coefficient at the driving wavelength~\cite{coen_convection_1999}. For more details concerning the normalization of~Eq.~\eqref{LLE}, see~\cite{leo_temporal_2010, hendry_spontaneous_2018}.

It is worth highlighting that, as noted in~\cite{parra-rivas_effects_2014}, there are numerous distinct mechanisms that can give rise to de-synchronization and drift. In addition to the pump pulse repetition rate being offset from the cavity FSR, drifts can arise due to higher-order linear and nonlinear effects (such as higher-order dispersion or stimulated Raman scattering) that shift the center wavelength of the CS away from the driving wavelength~\cite{milian_solitons_2015, yi_theory_2016, wang_stimulated_2018}. The drift term in Eq.~\eqref{LLE} can at least qualitatively capture the salient dynamics regardless of the physical origins of de-synchronization~\cite{parra-rivas_effects_2014}. Of course, if the higher-order linear or nonlinear effects are sufficiently strong --- such that their impact is not restricted to simply inducing a drift --- then Eq.~\eqref{LLE} must be augmented with additional terms that more rigorously describe the pertinent effects.

\section{Soliton trapping without de-synchronization}
We first recall CS trapping behaviour in the absence of de-synchronization ($d = 0$). In the presence of driving field amplitude inhomogeneities, a CS positioned at $\tau_\mathrm{CS}$ will (to first order) experience a drift in (fast) time with a rate~\cite{hendry_spontaneous_2018}:
\begin{equation}
    v_0 = \frac{d\tau_\mathrm{CS}}{dt} = a(S,\Delta) \left.\frac{dS}{d\tau}\right|_{\tau=\tau_\mathrm{CS}}.
    \label{driftv}
\end{equation}
Here, the proportionality coefficient $a(S,\Delta)$ corresponds to the projection of the CS's neutral (or Goldstone) mode along a linear fast time variation, and it depends on the local value of the driving field $S = S(\tau_\mathrm{CS})$ as well as the cavity detuning $\Delta$~\cite{maggipinto_cavity_2000,javaloyes_dynamics_2016}. From Eq.~\eqref{driftv}, it follows that the CS equilibrium (trapping) positions are given either by the positions where $dS/d\tau = 0$ or where the local driving field $S$ attains a critical value $S_\mathrm{c}$ for which $a(S_\mathrm{c},\Delta) = 0$. If the peak amplitude of the driving pulses satisfy $S_0 < S_\mathrm{c}$, the former ($dS/d\tau = 0$) trapping position is stable and the latter [$a(S_\mathrm{c},\Delta) = 0$] trapping positions do not exist. As the peak amplitude $S_0$ increases past the critical value $S_\mathrm{c}$, the $dS/d\tau = 0$ trapping position becomes unstable through a pitchfork bifurcation, concomitant with the emergence of a pair of new asymmetric stable states that correspond to the $a(S_\mathrm{c},\Delta) = 0$ equilibria~\cite{hendry_spontaneous_2018}. In this latter regime (where $S_0 > S_\mathrm{c}$), a CS will be attracted to (and trapped at) positions $\tau_\mathrm{c}$ on the edge of the driving pulse, where $S(\tau_\mathrm{c}) = S_\mathrm{c}$.

In Ref.~\cite{hendry_spontaneous_2018} it was shown that, for increasing detuning $\Delta$, the critical driving level $S_\mathrm{c}$ approaches the minimum driving level required for soliton existence [\mbox{$S_\mathrm{min} \approx (8\Delta/\pi^2)^{1/2}$}]. As most (microresonator) experiments operate in the regime $\Delta\gg1$, where $S_\mathrm{c}$ is very close to $S_\mathrm{min}$, it is reasonable to conclude that $S_0 > S_\mathrm{c}$ in most experimental situations. We accordingly focus our attention on this scenario throughout our manuscript~\cite{Note1}.

	\begin{figure}[t]	
		\includegraphics[width=\linewidth]{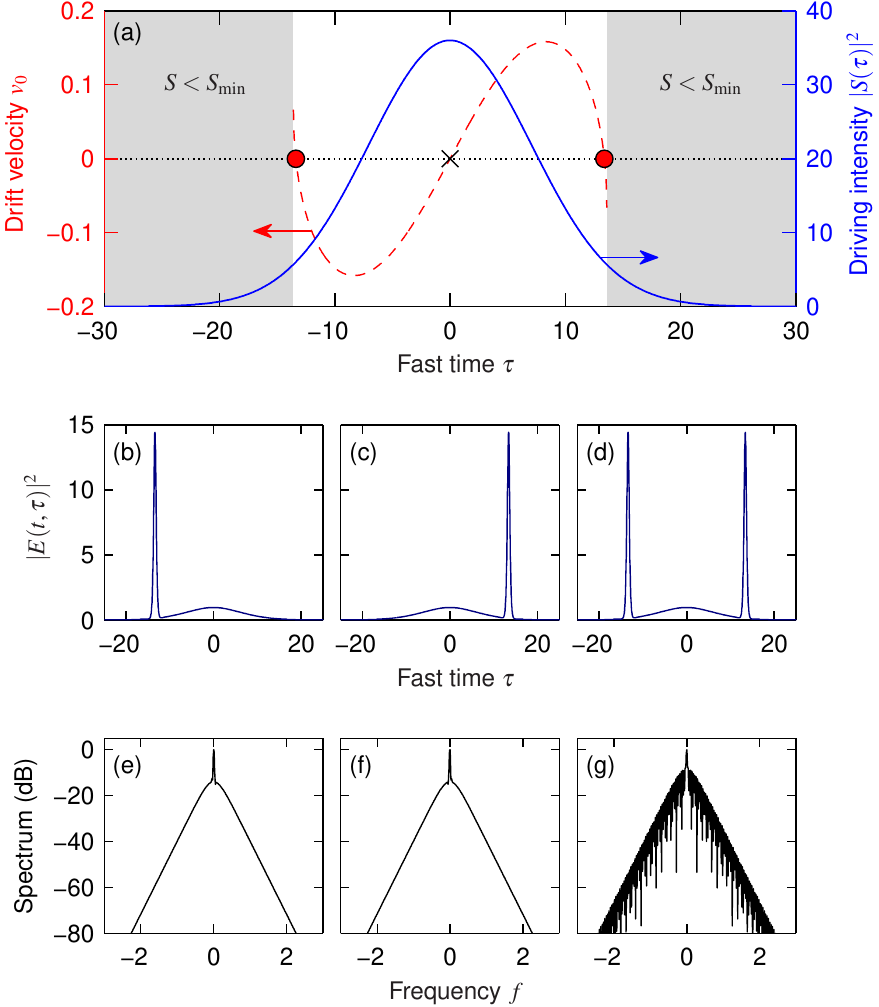}
		\caption{(a) Red dashed curve shows the drift velocity due to the amplitude inhomogeneity of a Gaussian driving field (blue solid curve) with amplitude $S_0 = 6$ and duration $\tau_\mathrm{g} = 10$. The cavity detuning $\Delta = 7$. Red solid circles (black cross) highlight stable (unstable) trapping positions. CSs do not exist in the gray shaded area, where the local driving field level is smaller than the minimum required for soliton existence. (b--d) Temporal profiles of the three possible soliton states that can manifest themselves for the parameters listed in (a). (e--g) Spectral profiles corresponding to (b--d), respectively.}
		\label{fig1}
\end{figure}

Figure~\ref{fig1}(a) depicts a typical drift velocity profile in the $S_0>S_\mathrm{c}$ regime, together with the intensity profile of the corresponding driving pulse [see caption for parameters]. As expected, there are two stable trapping positions $\tau_\mathrm{c}$ [i.e., positions where $S(\tau_\mathrm{c}) = S_\mathrm{c}$] that are symmetrically detuned with respect to the peak of the driving pulse. In steady-state, a CS can be trapped at either one of those positions, either singly or simultaneously. This is illustrated in Figs.~\ref{fig1}(b)--(g), which show examples of temporal [Figs.~\ref{fig1}(b--d)] and spectral [Figs.~\ref{fig1}(e--g)] profiles of numerically simulated steady state field configurations with no de-synchronization present. These results were obtained by numerically integrating~Eq.~\eqref{LLE} with a split-step Fourier algorithm, assuming initial conditions that result in the excitation of a CS slightly to the left [Fig.~\ref{fig1}(b, e)] or right [Fig.~\ref{fig1}(c, f)] of the peak of the driving field [Fig.~\ref{fig1}(d, g) was obtained by simultaneously exciting CSs at both sides of the peak].

It is interesting to note that the three configurations shown in Fig.~\ref{fig1} are the \emph{only} CS configurations that the chosen simulation parameters support. In particular, we find that excitation of multiple CSs on the same side of the pump pulse leads to merging or annihilation at the corresponding trapping position~\cite{jang_controlled_2016}. This behaviour appears to be universal: we have not identified parameters that would permit states with more than two CSs. We suspect, however, that the presence of strong higher-order effects can change the situation by giving rise to strongly bound soliton states~\cite{wang_universal_2017}.

\section{Impact of de-synchronization on CS existence and stability}
In the presence of de-synchronization, a CS will experience an additional convective drift at a rate that is governed by the drift coefficient $d$~\cite{parra-rivas_effects_2014}. The total drift of the CS relative to the pump pulse is then:
\begin{equation}
    v = v_0 + d = a(S,\Delta) \left.\frac{dS}{d\tau}\right|_{\tau=\tau_\mathrm{CS}}+d.
    \label{drift_full}
\end{equation}
Because of de-synchronization, the stable CS trapping sites are shifted to new positions where \mbox{$a(S,\Delta)dS/d\tau = -d$}; at these positions, drift due to de-synchronization is exactly balanced by the inhomogeneity of the driving field. It should be clear that, for typical pulsed driving fields (with $S_0>S_\mathrm{c}$), both trapping positions shift in the same direction, thus breaking the symmetry of the double-soliton state [cf. Fig.~\ref{fig1}(a)]. Compounded by the localized nature of a pulsed driving field, such shifting can influence the number of possible CS configurations. This is particularly evident at higher detunings (\mbox{$\Delta \gtrsim 5$}), where the trapping positions in the absence of de-synchronization are already very close to the minimum driving field value for which CSs can exist ($S_\mathrm{min}$)~\cite{hendry_spontaneous_2018}. In this regime, even a small de-synchronization can cause one of the trapping positions to fall below this minimum value, hence reducing the number of possible CS configurations from three to one.

	\begin{figure}[t]	
		\includegraphics[width=\linewidth]{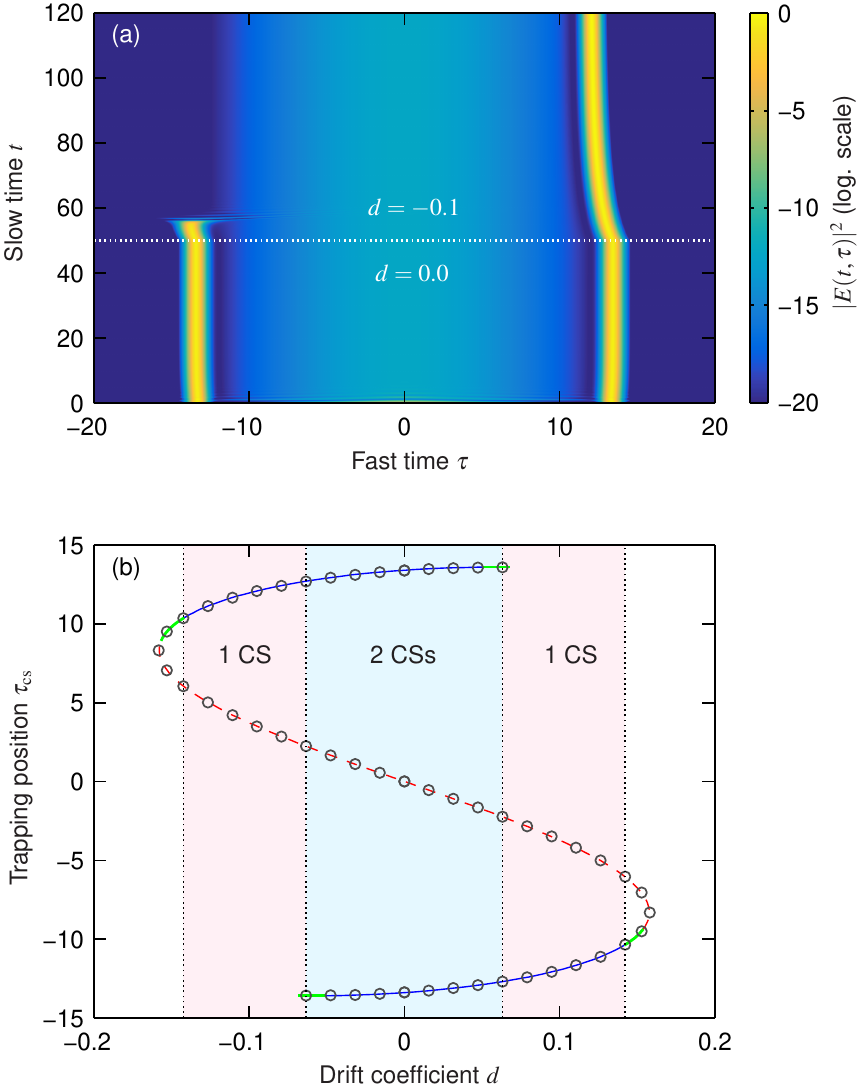}
		\caption{(a) False color plot showing the evolution of the intracavity field intensity when de-synchronization with \mbox{$d = -0.1$} is abruptly introduced at slow time $t = 50$ (highlighted by the horizontal dotted line). Other parameter as in Fig.~\ref{fig1}. (b) Solid blue, dashed red, and solid green curves show the positions of CS solutions of Eq.~\eqref{LLE} as a function of the drift coefficient $d$, obtained using a Newton continuation algorithm. The solid blue curve corresponds to stable solutions, dashed red curve corresponds to the trivially unstable $dS/d\tau = 0$ trapping position, while the solid green curves indicate more complex instabilities [see main text]. An additional branch of trivially unstable CSs that connects the two end-points is not shown for clarity. Open gray circles show trapping positions found by directly solving the roots of the equation $v = 0$, where $v$ is given by Eq.~\eqref{drift_full}.}
		\label{fig2}
\end{figure}

Figure~\ref{fig2}(a) shows results from numerical simulations that illustrate the dynamics described above [parameters as in Fig.~\ref{fig1}]. Here, de-synchronization is initially absent ($d = 0$) and two CSs are excited at either side of the pump pulse. At a slow time of $t = 50$, we introduce a de-synchronization ($d = -0.1$) which causes both CS trapping positions to shift towards the leading (left) edge of the pump pulse. Significantly, while the trailing (right) CS simply adjusts to its new trapping position, the leading CS ceases to exist as it is pushed below the minimum driving field amplitude $S_\mathrm{min}$.

A more comprehensive analysis of the impact of de-synchronization in the large-$\Delta$ regime is shown in Fig.~\ref{fig2}(b). Here, the solid blue (dashed red) curves show steady-state positions of stable (unstable) CSs solutions of Eq.~\eqref{LLE} as a function of the drift parameter $d$, obtained using a Newton continuation algorithm. Also shown, as open circles, are the trapping positions predicted by directly finding the roots of Eq.~\eqref{drift_full}. Several conclusions can be drawn. First, we see that the roots of Eq.~\eqref{drift_full} agree very well with the trapping positions extracted from Newton calculations. Second, for drift values close to zero, possible CS positions exhibit bistability (blue shaded area), which is indicative of the ability to sustain all three CS configurations demonstrated in Fig.~\ref{fig1}. Last, for values of $d$ outside this region of bistability, only one CS configuration, consisting of a single CS trapped either on the trailing ($d<0$) or leading ($d>0$) edge of the pump pulse, is possible (red shaded areas).

	\begin{figure}[t]	
		\includegraphics[width=\linewidth]{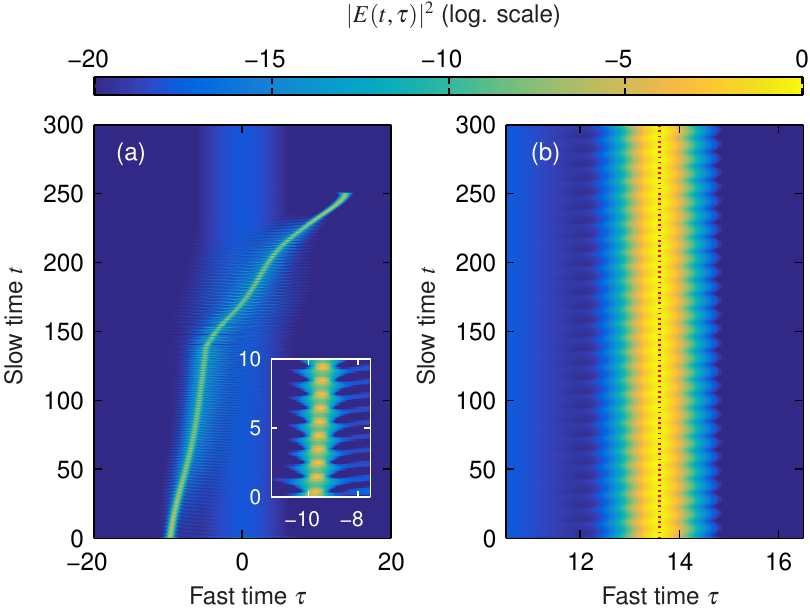}
		\caption{Simulated intracavity dynamics of unstable CSs for (a) $d = 0.15$ and (b) $d = 0.06$ with other parameters as in Fig.~\ref{fig1}. Both simulations use an initial conditions comprised of a soliton-like perturbation close to its predicted trapping position. The inset in (a) shows the evolution over a smaller range of slow time so as to highlight the oscillatory nature of the instability. The dotted vertical line in (b) indicates the temporal position $\tau_\mathrm{min}$ which satisfies $S(\tau_\mathrm{min}) = S_\mathrm{min}$. Note the different x-axes in (a) and (b).}
		\label{fig3}
\end{figure}

In addition to affecting the possible CS configurations, de-synchronization can also give rise to CS instabilities. In particular, for sufficiently large driving amplitudes, it is possible that one of the trapping positions shifts to a position where the local value of the driving field exceeds the well-known CS Hopf bifurcation threshold~\cite{leo_dynamics_2013}. In fact, this situation can be observed in Fig.~\ref{fig2}(b), where the CS solutions are found to become unstable for drift values $|d| > 0.14$ (solid green curves). Dynamical split-step simulations show that, in this regime, the CSs breathe with slow time $t$ (as in the case of CW driving~\cite{leo_dynamics_2013}), but do not remain trapped at a specific position. Rather, as shown in Fig.~\ref{fig3}(a), they become unlocked, pass over the peak of the driving pulse, and subsequently cease to exist. Interestingly, instabilities can also arise when de-synchronization shifts the CSs away from the peak of the driving pulse and closer to the minimum driving amplitude $S_\mathrm{min}$. More specifically, we find that, for a small range of de-synchronizations, a CS can exhibit persistent temporal oscillations around the point where their existence would be expected to cease under conditions of CW driving [see Fig.~\ref{fig3}(b)]. The range of de-synchronizations where such instabilities manifest themselves (as well as the magnitude of the temporal oscillations) appears to increase as the pump pulse duration decreases, presumably due to the corresponding increase in the underlying amplitude gradient.

The behaviour summarized in Fig.~\ref{fig2} manifests itself at higher detuning values ($\Delta \gtrsim 5$), as it is only in this regime that the trapping positions in the absence of de-synchronization are very close to the minimum value of CS existence~\cite{hendry_spontaneous_2018}. Notwithstanding, we have found that de-synchronization also affects CS existence at lower detunings, albeit through a slightly different mechanism. At these lower detunings, CS trapping positions in the absence of de-synchronization lie between the minimum and maximum values of CS existence. Accordingly, a CSs can tolerate a considerable shift \emph{away} from the peak of the driving pulse without ceasing to exist, in stark contrast to the larger detuning case discussed above. A situation can then arise where the trapping position shifting towards the peak of the pump pulse ceases to exist before the one shifting away from the peak. In this situation, if two CSs co-exist initially, de-synchronization forces the solitons to collide, leaving just one CS at the only remaining trapping position.

	\begin{figure}[t]	
		\includegraphics[width=\linewidth]{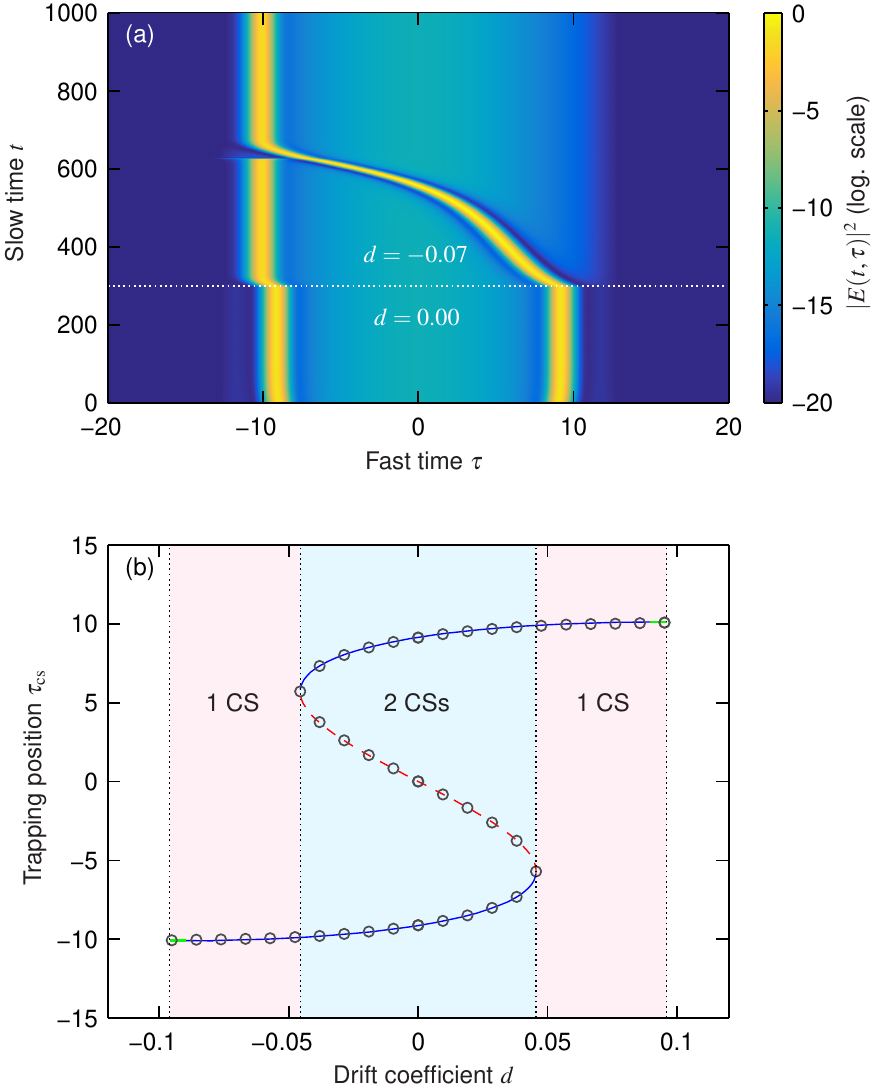}
		\caption{Numerical results for a cavity detuning of \mbox{$\Delta = 4$}, driving pulse amplitude $S_0 = 3$, and driving pulse width $\tau_\mathrm{g} = 10$. (a) Evolution of the intracavity field intensity when de-synchronization with $d = -0.07$ is abruptly introduced at slow time $t = 300$ (highlighted by the horizontal dotted curve). (b) As in Fig.~\ref{fig2}, blue, green, and red curves show the positions of stable and unstable CS solutions of Eq.~\eqref{LLE} as a function of the drift coefficient $d$. Open gray circles show trapping positions found by directly solving the roots of the equation $v = 0$, where $v$ is given by Eq.~\eqref{drift_full}.}
		\label{fig4}
\end{figure}

Figure~\ref{fig4}(a) shows simulation results that illustrate these dynamics [see caption for parameters]. The simulation starts from a two-soliton configuration in the absence of de-synchronization. At a slow time of $t = 300$, we introduce a small desynchronization ($d = -0.07$) and observe how the trailing CS becomes unlocked, traveling over the peak of the pump pulse towards the one remaining trapping position. As the leading CS is already occupying this position, the two CSs interact and merge together. In Fig.~\ref{fig4}(b), we plot the CS trapping positions as a function of the drift coefficient for $\Delta  = 4$ [curves obtained in the same fashion as those shown in Fig.~\ref{fig2}(b)]. We again see the S-shape characteristic to bistability; however, in contrast to the large-detuning case [cf. Fig.~\ref{fig2}(b)], the sign of the drift coefficient required to ensure a single trailing or leading CS is the opposite. This is simply a manifestation of the different mechanism that underpins the removal of the two-soliton state.

The results reported above show that pump de-synchronization can be harnessed to ensure single-soliton operation in Kerr resonators~\cite{cole_kerr_2018}. Of course, it should be clear that the de-synchronization has to be sufficiently small [within the shaded regions of Fig.~\ref{fig2}(b) and Fig.~\ref{fig4}(b)] so that one of the trapping positions persists, allowing a CS to remain (frequency) locked to the driving field. The locking range, i.e., the range of de-synchronization that a CS can tolerate, can be obtained from Eq.~\eqref{drift_full}. Specifically, the maximum desynchronization that can be compensated for by the pump inhomogeneity is:
\begin{equation}
     |d_\mathrm{max}|=\mathrm{max}\left|a(S,\Delta) \left.\frac{dS}{d\tau}\right.\right|.
    \label{driftv2}
\end{equation}
\noindent
In dimensional units, this yields a maximum tolerable drift per round trip of $\Delta t_\mathrm{max} = d_\mathrm{max}\sqrt{|\beta_2|L\pi/(2\mathcal{F})}$ and a corresponding repetition frequency mismatch of $\Delta f_\mathrm{max} = \mathrm{FSR}^{2}\Delta t_\mathrm{max}$. Because the coefficient $a(S,\Delta)$ depends on the local value of the driving field, it is not possible to express the locking range as a simple product involving a constant coefficient and the maximum slope of the driving field (as is the case for phase modulated driving fields~\cite{jang_temporal_2015}). Rather, evaluating the locking range requires knowledge of the full functional dependence of $a(S,\Delta)$.

\section{Detuning-dependent dynamics and effect of stimulated Raman scattering}
Our analysis has so far revolved around isolated values of the cavity detuning. To gain more insights, we next consider how de-synchronization affects the cavity dynamics as the detuning is continuously (and adiabatically) scanned over a resonance. To facilitate comparisons with prior experiments, we consider a comparatively short driving pulse with $\tau_\mathrm{g} = 4$; this value coincides with the pulse duration used in the experiments reported in~\cite{obrzud_temporal_2017}. We perform simulations for a wide range of different drift coefficients $d$, and for each value, we simulate the intracavity dynamics as the detuning is slowly scanned from negative to positive values.

	\begin{figure}[t]	
		\includegraphics[width=\linewidth]{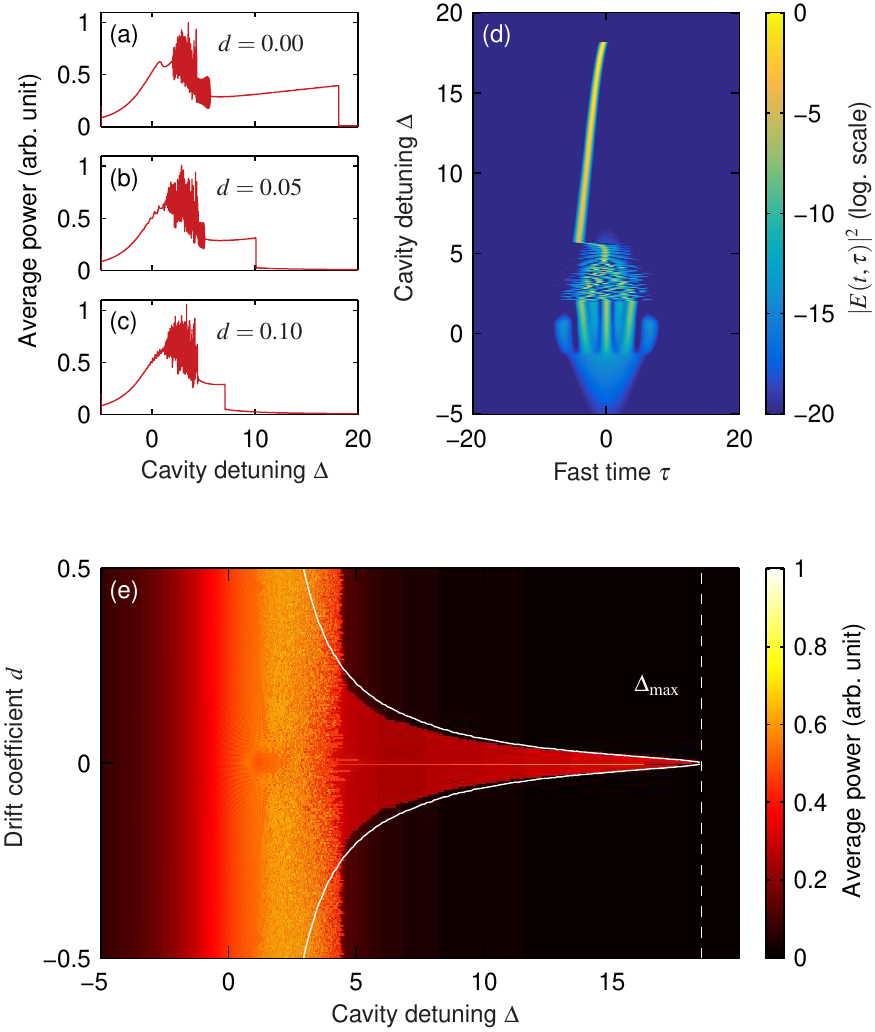}
		\caption{(a--c) Average intracavity power as the detuning is scanned over the cavity resonance for different drift coefficients $d$ as indicated. (d) Dynamical evolution of the intracavity field intensity as the detuning is scanned over the resonance for $d = 0$. (e) False colour plot of scan traces as in (a--c) for a range of different drift coefficients $d$. Solid white curves highlight the maximum de-synchronization that a CS can tolerate when excited by scanning the detuning [as described in the main text]. The dashed vertical white line corresponds to the absolute maximum detuning of CS existence, i.e., $\Delta_\mathrm{max} = \pi^2S_0^2/8$. Simulations use $S_0 = \sqrt{15}$ and $\tau_\mathrm{g} = 4$.}
		\label{fig5}
\end{figure}

Figures~\ref{fig5}(a)--(c) depict the evolution of the average intracavity power as the detuning is scanned for three different values of the drift coefficient $d$ [see caption]. For each value, the evolution of the intracavity power is indicative of well-known Kerr cavity dynamics: a chaotic modulation instability regime is followed by a distinct ``soliton step''~\cite{herr_temporal_2014}. The length of the soliton step is found to decrease with increasing magnitude of the drift coefficient $|d|$ --- a feature that will be discussed in the following paragraph. Figure~\ref{fig5}(d) shows temporal dynamics for the $d = 0$ case, and we see how, at a detuning of about $\Delta \approx 5$, a single CS forms on the leading edge of the driving pulse, persisting until the detuning reaches the maximum value of CS existence, $\Delta_\mathrm{max} = \pi^2S_0^2/8 \approx 18.5$. It is worth noting that, because for $d = 0$ the system is perfectly symmetric, a CS can be expected to form with equal probability on the leading or the trailing edge of the driving pulse. In contrast, for $d\neq0$, the symmetry of the system is explicitly broken~\cite{garbin_asymmetry_2019}, and the solitons preferentially form either on the leading ($d<0$) or the trailing ($d>0$) edge of the pump pulse.

Figure~\ref{fig5}(e) shows the evolution of the average intracavity power over a much wider range of drift coefficients. We can make two important observations. First, the soliton portions of the scan traces are (approximately) symmetric with respect to $d = 0$ (as expected based on the symmetry of the driving pulse). Second, the range of soliton existence (in detuning) decreases as the magnitude of the drift coefficient increases. This latter observation can be explained as follows. As detuning increases, the critical driving field value $S_\mathrm{c}$ towards which the solitons are attracted in the absence of de-synchronization ($d = 0$) asymptotically approaches $S_\mathrm{min}$ --- the minimum value for which solitons can still exist~\cite{Note2}. For $d > 0$ ($d < 0$), the explicit symmetry breaking induced by the de-synchronization favours soliton formation on the trailing (leading) edge of the driving pulse, which is also the direction towards which the CS trapping positions are shifted. Taken together, as the detuning increases, the de-synchronization needed to push the CS below the minimum driving level $S_\mathrm{min}$ decreases. Accordingly, the range of detunings over which CSs can exist decreases as the magnitude of de-synchronization increases.

To corroborate our explanation, the white curves in Fig.~\ref{fig5}(e) highlight the maximum de-synchronization that a CS on the trailing or leading edge of the pump pulse can tolerate when $d>0$ or $d<0$, respectively, as a function of detuning $\Delta$. These curves were obtained by evaluating the velocity $v = a(S_\mathrm{min},\Delta)dS/d\tau|_{\tau=\tau_\mathrm{min}}$, where $\tau_\mathrm{min}$ satisfies $S(\tau_\mathrm{min}) = S_\mathrm{min}$. As can be seen, the theoretical predictions are in excellent agreement with the range of soliton existence observed in numerical simulations. It is worth emphasizing that the maximum tolerable de-synchronization depicted in Fig.~\ref{fig5}(e) does not coincide with $d_\mathrm{max}$ given by Eq.~\eqref{driftv2}: the latter represents an absolute maximum, and may require that the CS resides on the leading (trailing) edge of the driving pulse when $d>0$ ($d<0$), which is not the case when the solitons form spontaneously as the detuning is scanned.

It is very interesting to note that the simulation results shown in Fig.~\ref{fig5}(e) differ markedly from corresponding experimental results reported in Refs.~\cite{obrzud_temporal_2017}. In particular, experiments show soliton steps that are distinctly asymmetric as a function of de-synchronization. We suspect this discrepancy arises from the presence of some other mechanism that explicitly breaks the system's symmetry, hence affecting the soliton trapping positions. Recalling that the resonator used in Ref.~\cite{obrzud_temporal_2017} was made out of fused silica, we may identify stimulated Raman scattering (SRS) as a potential process that could explain the observations.

To test our hypothesis, we have repeated the simulations shown in Fig.~\ref{fig5}(d) with the inclusion of the full Raman response of fused silica. More specifically, our simulations use the following modified LLE~\cite{wang_stimulated_2018}:
	\begin{equation}
	\small
	\begin{split}
	\frac{\partial E(t,\tau)}{\partial t}=&\left[ -1-i\Delta-d\frac{\partial}{\partial\tau}+i\frac{\partial^2}{\partial \tau^2}\right]E+S(\tau)\\&+i\Big[(1-f_\mathrm{R})|E|^2+f_\mathrm{R}\left[\Gamma(\tau,\tau_\mathrm{s})*|E|^2\right]\Big]E.
	\end{split}
	\label{LEE2}
	\end{equation}
Here, $f_\mathrm{R}$ is the fraction of the instantaneous nonlinearity that is due to SRS and $\Gamma(\tau,\tau_\mathrm{s})$ is a normalized response function that is related to the usual Raman response function $h_\mathrm{R}(\tau)$ through $\Gamma(\tau,\tau_\mathrm{s})=\tau_\mathrm{s}h_\mathrm{R}(\tau\tau_\mathrm{s})$, where the normalization timescale $\tau_\mathrm{s}=\sqrt{\mathcal{F}|\beta_2|L/(2\pi)}$. Modelling the silica fiber Fabry-Perot resonator of Ref.~\cite{obrzud_temporal_2017}, we have $f_\mathrm{R} = 0.18$,  $\tau_\mathrm{s} = 300~\mathrm{fs}$, and we use the well-known multiple-vibrational mode model to evaluate the response function $h_\mathrm{R}(\tau)$~\cite{Hollenbeck}.

	\begin{figure}[t]	
		\includegraphics[width=\linewidth]{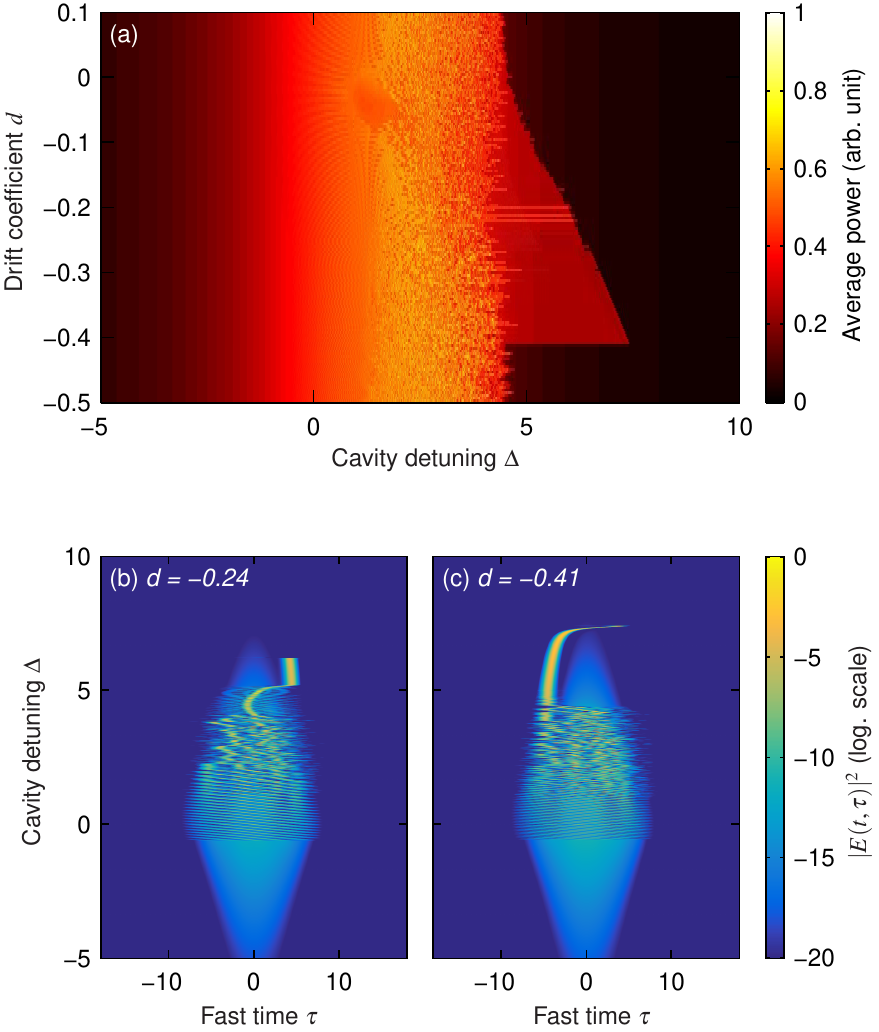}
		\caption{(a) Soliton steps as a function of the drift coefficient $d$ when SRS is included in the model. Note the different axes limits compared to Fig.~\ref{fig5}(e). (b, c) Dynamical evolutions of the intracavity field intensity for drift coefficients as indicated. The Raman response function used in the simulations corresponds to fused silica~\cite{Hollenbeck}, with $f_\mathrm{R} = 0.18$. Other parameters as in Fig.~\ref{fig5}.}
		\label{fig6}
\end{figure}

Figure~\ref{fig6}(a) shows numerically simulated soliton steps as a function of the drift coefficient $d$ in the presence of SRS. Comparing the results with those obtained in the absence of SRS [cf. Fig.~\ref{fig5}(d)], we immediately notice two important differences. First, in accordance with corresponding experimental findings~\cite{obrzud_temporal_2017}, we now find the steps to be asymmetric with $d$. Second, we observe that the soliton steps manifest themselves only when $d<0$. These observations can be understood by recalling that the redshift induced by SRS gives rise to a time-domain drift~\cite{wang_stimulated_2018,milian_solitons_2015} that shifts the CS trapping positions towards the trailing edge of the driving pulse. For the parameters under study, the effect is so strong that a negative de-synchronization is required to compensate for the SRS-induced drift and to push the trapping positions above the minimum level of soliton existence ($S_\mathrm{min}$). This competition between SRS and de-synchronization also explains why the range of soliton existence initially increases as the magnitude of the drift coefficient increases: a larger drift pushes the trapping positions further away from the minimum level of soliton existence, thus allowing the solitons to withstand the opposing effect of SRS over a wider range of detunings. (Note in this context that the SRS-induced soliton redshift is known to grow stronger with increasing detuning~\cite{wang_stimulated_2018}.) It is worth noting that the interplay between SRS and de-synchronization can be seen as an example of \emph{asymmetric balance}, whereby two distinct mechanisms of explicit symmetry breaking (here SRS and de-synchronization) compete and compensate for each other~\cite{garbin_asymmetry_2019}.

Another conspicuous effect of SRS is the overall reduction in the range of soliton existence. Indeed, whilst solitons are found to exist up to $\Delta\approx 18.5$ in the absence of SRS [cf. Fig.~\ref{fig5}(d)], they cease to exist already at $\Delta \approx 7.5$ in the presence of SRS (under optimal de-synchronization). This behaviour agrees with recent research that shows SRS to limit the range of CS existence~\cite{wang_stimulated_2018}. Figures~\ref{fig6}(b) and (c) show typical evolutions of the intracavity field intensity for two different values of the drift coefficient. When the magnitude of the drift coefficient $d$ is small, the CSs are formed on the trailing edge of the driving pulse due to SRS being the dominant mechanism of explicit symmetry breaking. As the magnitude of the drift coefficient increases, this ceases to be the case, and the solitons begin to predominantly form on the leading edge of the pulse. In both cases, we see complex dynamics that eventually lead to the soliton ceasing to exist.

The results shown in Fig.~\ref{fig6}(a) are in reasonable qualitative agreement with experimental results in~\cite{obrzud_temporal_2017}. In dimensional units corresponding to those experiments, our simulations predict that the CSs can exist over a 27~kHz range of de-synchronization. While smaller than the 100~kHz observed in experiments, we attribute the discrepancy to differences in the driving pulse characteristics (e.g. peak power and phase profile) as well as the rate with which the detuning is changed. On the other hand, the (reduced) range of detunings over which the CSs exist in our simulations appears to be in very good agreement with the range observed experimentally. Taken together, we believe that the interplay between pump-cavity de-synchronization and SRS underpins the asymmetric soliton steps observed in Ref.~\cite{obrzud_temporal_2017}. Of course, we must emphasize that other higher-order effects that give rise to CS drift --- such as e.g. higher-order dispersion or self-steepening --- are also expected to give rise to asymmetric soliton steps, provided these effects are sufficiently strong. For the silica resonator used in~\cite{obrzud_temporal_2017}, it is however reasonable to assume SRS to be the dominant perturbation.

\section{Conclusions}
In conclusion, we have performed a systematic numerical and theoretical investigation of the effects of de-synchronization and drift on soliton Kerr frequency comb generation in the presence of pulsed driving fields. Our results show that de-synchronization can impact the positioning, stability, and existence of CSs. In particular, we have shown that de-synchronization can be leveraged to ensure the generation of single-soliton states via two different mechanisms depending on the cavity detuning. We have also studied the interplay between SRS and pump-cavity de-synchronization, obtaining strong evidence that asymmetric soliton steps observed in recent experiments arise precisely due to such interplay. Our findings shed light on the dynamics of CSs in the presence of pulsed driving fields, and could have impact on the design of efficient microresonator frequency combs~\cite{obrzud_temporal_2017,obrzud_microphotonic_2019} or free-space enhancement cavities~\cite{lilienfein_temporal_2019}.

\section*{Acknowledgments}
The authors wish to acknowledge the Centre for eResearch at the University of Auckland for their help in facilitating this research. We also acknowledge financial support from the Marsden Fund, the Rutherford Discovery Fellowships, and the James Cook fellowships of the Royal Society of New Zealand.

\end{document}